\documentclass[prl,reprint,superscriptaddress,preprintnumbers]{revtex4-2}
\pdfoutput=1
\usepackage{amsfonts, amssymb, amsmath}
\usepackage{mathrsfs}
\usepackage[utf8]{inputenc}
\usepackage[russian,english]{babel}

\usepackage{color}
\definecolor{dark-gray}{gray}{0.20}
\definecolor{gray}{gray}{0.30}
\definecolor{light-gray}{gray}{0.80}
\definecolor{dark-red}{rgb}{0.7,0,0}
\definecolor{dark-green}{rgb}{0.1,0.4,0}
\definecolor{dark-blue}{rgb}{0.3,0.3,0.7}
\definecolor{light-blue}{rgb}{0.8,0.8,1}
\definecolor{blue}{rgb}{0,0,1}
\definecolor{red}{rgb}{1,0,0}
\definecolor{green}{rgb}{0,1,0}

\usepackage{hyperref}
\hypersetup{
	colorlinks=true,
	linkcolor=dark-blue,
	citecolor=dark-red,
	urlcolor=dark-blue,
	linktoc=page
}

\def\i{{\rm i}}

\newcommand{\be}{\begin{equation}}
\newcommand{\ee}{\end{equation}}
\newcommand{\bea}{\begin{eqnarray}}
\newcommand{\eea}{\end{eqnarray}}

\begin{document}

\title{Explicit black hole thermodynamics in natural variables}

\author{{\small an observation by}\\ Kiril Hristov}

\affiliation{Faculty of Physics, Sofia University, J. Bourchier Blvd. 5, 1164 Sofia, Bulgaria}
\affiliation{INRNE, Bulgarian Academy of Sciences, Tsarigradsko Chaussee 72, 1784 Sofia, Bulgaria}

\begin{abstract}
We consider the general thermal asymptotically flat Kerr-Newman black holes in 4d Einstein-Maxwell theory. Even though their thermodynamics has been understood for decades, the Gibbs free energy and on-shell action are only known implicitly as functions of the standard chemical potentials. Using the so-called left and right moving (or holomorphic and anti-holomorphic) variables related to the chemical potentials on both the outer and the inner horizons, we are able to present explicit and very simple expressions for all quantities. We discuss various limits in the parameter space, remarkably finding a smooth BPS limit allowing direct access to the extremal surface. In the BPS limit the anti-holomorphic part of the on-shell action vanishes identically, leading automatically to the holomorphic expression expected microscopically. This gives us confidence that the newly defined thermal partition function in terms of these variables is the natural candidate for a full microscopic description.
\end{abstract}
\date{\today}
\maketitle


\subsection{Introduction and main results}
\label{sec:intro}
\vspace{-3mm}

The study of black hole thermodynamics was initiated in the 70's by some notable works such as \cite{Bekenstein:1973ur,Hawking:1975vcx,Gibbons:1976ue}, and is presently a well-explored subject that still remains of large interest and at times heated debate. The underlying microscopic description is one of the great open questions and motivations for quantum gravity. In the current work we revisit the semi-classical regime of 4d Kerr-Newman thermodynamics and focus on a problem that might at first seem rather artificial. The black hole entropy, being proportional to the area, in the microcanonical ensemble can be simply expressed in terms of the asymptotic charges: the mass $M$, angular momentum $J$ and electromagnetic charges $Q$ and $P$. However, without imposing additional boundary conditions (see below), the on-shell action \cite{Gibbons:1976ue,York:1986it,Braden:1990hw} is proportional to the Gibbs free energy in the grand-canonical ensemble,~\footnote{Special care is needed when applying these words to magnetic charges, see \cite{Hawking:1995ap}. We give more precise statements in the bulk of this paper.}, thus it is a function of the corresponding inverse temperature $\beta = 1/T$, angular velocity $\Omega$ and electromagnetic potentials $\Phi$ and $\Psi$, conjugate variables to the conserved charges in the same order. Writing all expressions as functions of the chemical potentials is however very unwieldy as it requires solutions for higher roots. The usual way around is to numerically plot the quantities of thermodynamic interest and understand the qualitative features pictorially.

Another point of view, which we advocate here, is that $\beta, \Omega, \Phi, \Psi$ are {\it not} the natural set of variables in the grand-canonical potential. We instead argue in favor of the so-called left and right moving (or holomorphic and anti-holomorphic) variables  known from the 90's, see \cite{Cvetic:1997uw,Cvetic:1997xv,Wu:2004yk}. First let us emphasize the fact that the thermal black holes exhibit two event horizons, known as the outer (at radius $r_+$) and inner one (at radius $r_-$), see \eqref{eq:radii}. One can define consistent thermodynamic relations independently on {\it both} horizons, \cite{1979NCimB..51..262C}, with chemical potentials $\beta_+, \Omega_+, \Phi_+, \Psi_+$ and $\beta_-, \Omega_-, \Phi_-, \Psi_-$ and entropies $S_+, S_-$ defined at $r_+$ and $r_-$, respectively, see \eqref{eq:beta}-\eqref{eq:entropy} and \cite{Cvetic:2018dqf} for review. The respective potentials $I_+$ and $I_-$, obtained by integrating the on-shell action from the respective horizon, obey the quantum statistical relation,~\footnote{For the moment we set $P= 0$ leading to the conjugate variables $\Psi_\pm = 0$, and we reintroduce them later as their appearance is fixed by symmetry.}
\be
\label{eq:qsr}
	I_\pm (\beta_\pm, \Omega_\pm, \Phi_\pm) = \beta_{\pm} M - S_\pm - \beta_\pm \Omega_{\pm} J - \beta_{\pm} \Phi_{\pm} Q\ .
\ee
The meaning of the inner horizon thermodynamics can be disputed as temperature and entropy are no longer manifestly positive, but in fact the Euclideanized background of the Kerr-Newman black hole can be defined for a general set of {\it complex} asymptotic charges and conjugate potentials where $r_+ \ngtr r_-$, see \cite{Brown:1990fk}. In the present work we have the privilege to remain agnostic on this issue, but use our freedom to explore the full complex parameter space including the BPS limit.

We solve the ``explicitness'' problem in the following way. Defining the left and right moving variables
\bea
\label{eq:newvar}
\begin{split}
	\beta_{l,r} :=& \frac12\, (\beta_+ \pm \beta_-)\ , \quad \omega_{l,r} :=  \frac12\, (\beta_+ \Omega_+ \pm \beta_- \Omega_-)\ , \\  \varphi_{l,r} :=& \frac12\, (\beta_+ \Phi_+ \pm \beta_- \Phi_-)\ , \, \xi_{l,r} := \frac12\, (\beta_+ \Psi_+ \pm \beta_- \Psi_-)\ ,
\end{split}
\eea
and the respective entropies and on-shell actions,~\footnote{The left and right moving chemical potentials and entropies were in some form defined and analyzed in \cite{Cvetic:1997uw,Cvetic:1997xv}, and as explained in due course often used in  \cite{Cvetic:2010mn,Castro:2012av} and references thereof. The authors of these works however did not proceed further with the definition of the corresponding on-shell actions and their variations, and thus all explicit formulae we present below appear for the first time here.}
\be
\label{eq:newvar2}
	S_{l,r} :=  \frac12\, (S_+ \pm S_-)\ , \quad
	I_{l,r} := \frac12\, ( I_+ \pm I_-)\ ,
\ee
we show that
\be
\label{eq:qsrnew}
	I_{l,r} (\beta_{l,r}, \omega_{l,r}, \varphi_{l,r}) = \beta_{l,r} M - S_{l,r} - \omega_{l,r} J - \varphi_{l,r} Q\ ,
\ee
satisfy the first law of thermodynamics, $\delta I_{l,r} = 0$.  Furthermore, we explicitly find $\omega_l = 0$ and, ~\footnote{There are several branches of solutions whenever the inversion of the potentials requires square roots. Here we present the positive determination or the {\it principal} choice for all square roots, corresponding in particular to positive real values of the quantities inside the square roots. It is straightforward to take a different branch and change the corresponding signs in the explicit formulae we present.},
\bea
\begin{split}
\label{eq:expl}
		I_l (\beta_l, \varphi_l) = & \frac1{8 \pi}\, (\beta_l^2 - 2 \varphi_l^2)\ , \\
	I_r (\beta_r, \omega_r, \varphi_r) = \tfrac18\, & (3 \beta_r - \sqrt{\beta_r^2 + 8 \varphi_r^2}) \\ & \times  \sqrt{\frac{\beta_r^2 - 4 \varphi_r^2 + \beta_r \sqrt{\beta_r^2 + 8 \varphi_r^2}}{2\, (4 \pi^2 + \omega_r^2)}} \ ,
\end{split}
\eea
such that the thermodynamics of the left moving sector is effectively {\it static}. The original on-shell actions $I_\pm$ can be written as functions of the left and right moving variables,
\be
	I_\pm (\beta_{l,r}, \omega_r, \varphi_{l,r}) = I_l (\beta_l, \varphi_l) \pm I_r (\beta_r, \omega_r, \varphi_r)\ .
\ee
Due to electromagnetic duality, we can restore the magnetic charge $P$ and its conjugate potentials $\xi_{l,r}$ by replacing $Q^2$ with $Q^2 + P^2$ and $\varphi_{l, r}^2$ with $\varphi_{l,r}^2 + \xi_{l,r}^2$ in the grand-canonical ensemble. We comment on the mixed ensemble of varying $Q$ and fixed $P$ below.

These results lead us to a remarkable observation: upon taking the so-called BPS limit, $M = \sqrt{Q^2 + P^2}$, we can show that $I_l = (\varphi_l^2+\xi_l^2)/4 \pi$ and $I_r = 0$, i.e.\ the on-shell action is purely holomorphic. This is important as it precisely reproduces the microscopic expectations, available in this limit via string theory embeddings, see \cite{Strominger:1996sh,Maldacena:1997de},~\footnote{Strictly speaking, all string theory embedding require further Maxwell and scalar fields, i.e.\ {\it non-minimal} supergravity. Qualitatively, the present results carry through in these cases as well, \cite{Hristovtoapp}.}. Notice that there is an alternative approach to the BPS limit, detailed in the present case in \cite{Hristov:2022pmo} in analogy to the AdS results of \cite{Cassani:2019mms}. This approach uses rather different variables, which seem unable to solve the explicitness problem in the general thermal case. 

Finally, let us comment on the potential deeper meaning of these results. Even though at a semi-classical level all we have done is to rewrite the well-known laws of thermodynamics in different variables, at a full quantum level this allows us to define a genuinely new thermal partition function, see \eqref{eq:newpf}. It is identical with the standard thermal partition function at leading order, but suggests a priori different path integral and might prove important in the quest of determining the quantum corrections in a UV finite microscopic theory.

\vspace{-5mm}
\subsection{Kerr-Newman black holes}
\label{sec:setup}
\vspace{-3mm}

We consider Einstein-Maxwell theory with Newton constant set to unity, $G_N = 1$,
\be
	 I = \frac1{16 \pi} \int_{\cal M}  {\rm d}^4 x\, \sqrt{-g}\, \left( R - \frac14 F_{\mu \nu} F^{\mu \nu} \right)\ ,
\ee
with an abelian field strength $F = {\rm d} A$ and mostly positive metric signature. The most general stationary black hole solution in this theory is given by the line element
\bea
\begin{split}
	{\rm d} s^2 = &- \frac{\Delta (r)}{\rho^2} \left( {\rm d} t - a \sin \theta\, {\rm d} \phi \right)^2 + \frac{\rho^2}{\Delta (r)}\, {\rm d} r^2 \\ 
&+ \rho^2\, {\rm d} \theta^2 +  \frac{\sin^2 \theta}{\rho^2} \left( a\, {\rm d} t - (r^2+a^2)\, {\rm d} \phi \right)^2\ ,
\end{split}
\eea
with 
\be
	\Delta(r) = r^2 - 2 M r + a^2 + Q^2 + P^2\ , \, \, \, \rho^2 = r^2 + a^2 \cos^2 \theta\ ,
\ee
and background gauge field
\be
	A = - \frac{Q\, r}{\rho^2}\,  \left( {\rm d} t - a \sin \theta\, {\rm d} \phi \right) -  \frac{P\, \cos \theta}{\rho^2}\,   \left( a\, {\rm d} t - (r^2+a^2)\, {\rm d} \phi \right)\ ,
\ee
leading to the conserved electromagnetic charges $Q$ and $P$. The full solution is completely specified by $Q, P$, the mass (or energy) $M$ and the angular momentum $J = a M$. The two roots of the function $\Delta (r)$ are given by
\be
\label{eq:radii}
	r_{\pm} = M \pm \sqrt{M^2 - a^2 - Q^2 - P^2}\ ,
\ee
and correspond to the positions of the outer and the inner event horizons. We remind the reader that the theory enjoys electromagnetic duality such that all formulae presented at $P = 0$ can be uniquely extended to an arbitrary value of $P$. 

\vspace{-5mm}
\subsection{Old and new thermodynamic potentials}
\label{sec:thermo}
\vspace{-3mm}
In order to define the laws of black hole thermodynamics on each horizon, let us first note that the existence of two different horizons allows us to impose two different global regularity conditions on the Euclidean metric and gauge fields. Imposing regularity at the outer horizon, e.g.\ by requiring that there are no conical singularities, fixes the value of the intensive quantities such as temperature, angular velocity and electromagnetic potentials to a value with a ``$+$'' subscript. Imposing regularity on the inner horizon instead leads to these quantities with a ``$-$'' subscript. Notice that it is {\it impossible} to ensure global regularity on both horizons at the same time, such that the ``$+$'' and ``$-$'' thermodynamic laws we discuss are independent of each other. See \cite{Cvetic:2018dqf} for a physical interpretation of what observers feel in relation to their radial position, and for more careful definition of all these standard black hole quantities. 

Following the above prescription, the inverse temperature (given by the periodicity of the Wick rotated time coordinate) and angular velocity at the two horizons read
\be
\label{eq:beta}
	\beta_\pm = 2 \pi \frac{r_\pm^2 + a^2}{r_\pm - M}\ , \quad \quad \Omega_\pm = \frac{a}{r_\pm^2 + a^2}\ ,
\ee
while the electromagnetic potentials correspond to the horizon values of the temporal part of the gauge field and its magnetic dual,
\be
	\Phi_\pm = \frac{Q\, r_\pm}{r_\pm^2 + a^2}\ , \quad \quad \Psi_\pm = \frac{P\, r_\pm}{r_\pm^2 + a^2}\ .
\ee
The Bekenstein-Hawking entropies are given
\be
\label{eq:entropy}
	S_\pm = \pi (r_\pm^2 + a^2)\ ,
\ee
allowing one to explicitly verify the first law of black hole thermodynamics,
\be
\label{eq:firstlaw}
	\beta_\pm\, \delta M = \delta S_\pm + \beta_\pm \Omega_\pm\, \delta J + \beta_\pm \Phi_\pm\, \delta Q + \beta_\pm \Psi_\pm\, \delta P\ .
\ee
For the evaluation of the on-shell action we need the Gibbons-Hawking-York boundary term, \cite{Gibbons:1976ue,York:1986it}, 
\be
	I_\text{bdy} = - \frac1{8 \pi} \int_{\partial {\cal M}} {\rm d}^3 x\, \sqrt{\gamma} \left( K - K^0 \right)\ ,
\ee
where $\gamma$ is the induced metric on a constant radial slice, $K$ the extrinsic curvature and $K^0$ the constant boundary term of empty flat space. \eqref{eq:qsr} follows from $I_\pm := I |_{r_\pm}^\infty + I_\text{bdy}^\infty$ in Euclidean signature ($\tau := \i\, t$) with the indicated radial bounds and $\tau \in [0, \beta_\pm), \theta \in [0, \pi], \phi \in [0, 2 \pi)$, see \cite{Brown:1990fk,Brown:1990di} for details. In presence of $P$, this calculation gives a mixed ensemble of fixed $P$, \cite{Hawking:1995ap},
\be
	\tilde I_\pm  (\beta_\pm, \Omega_\pm, \Phi_\pm, P)  =  \beta_{\pm} M - S_\pm - \beta_\pm \Omega_{\pm} J - \beta_{\pm} \Phi_{\pm} Q\ ,
\ee
while we reserve the notation $I_\pm$ for the full grand-canonical ensemble with $\Psi_\pm$. Due to the form of $r_\pm$ involving a square root, an explicit attempt of finding $M, J, Q, P$ in terms of $\beta_\pm, \Omega_\pm, \Phi_\pm, \Psi_\pm$ does not lead to reasonable expressions and therefore $I_\pm$ and $\tilde I_\pm$ remain implicit functions of the chemical potentials, with the notable exception of two special limits discussed below.

Upon the introduction of the new thermodynamic potentials, defined in \eqref{eq:newvar}-\eqref{eq:newvar2}, the corresponding rearrangement of \eqref{eq:firstlaw} leads to
\be
\label{eq:lrfirstlaw}
	\beta_{l,r}\, \delta M = \delta S_{l,r} + \omega_{l,r}\, \delta J + \varphi_{l,r}\, \delta Q + \xi_{l,r}\, \delta P\ ,
\ee
such that the on-shell actions $I_{l,r}$ can be regarded as functions of the new variables. The new potentials can be then evaluated in terms of the asymptotic charges,
\bea
\label{eq:lrinasympt}
\begin{split}
	\beta_l &= 4 \pi\, M\ , \quad \quad \beta_r = 2 \pi\, \frac{2 M^2 - Q^2 - P^2}{\sqrt{M^2 - a^2 - Q^2 - P^2}} \ , \\
	\omega_l &= 0\ , \quad \quad \omega_r = \frac{2 \pi a}{\sqrt{M^2 - a^2 - Q^2 - P^2}}\ , \\
	\varphi_l & = 2 \pi Q\ , \quad \quad \varphi_r = \frac{2 \pi M Q}{\sqrt{M^2 - a^2 - Q^2 - P^2}}\ , \\
	\xi_l & = 2 \pi P\ , \quad \quad \xi_r = \frac{2 \pi M P}{\sqrt{M^2 - a^2 - Q^2 - P^2}}\ , \\
	S_l & = 2 \pi (2 M^2 - Q^2 - P^2) , \, S_r = 2 \pi M \sqrt{M^2 - a^2 - Q^2 - P^2} .
\end{split}
\eea
The above expressions are easily inverted to give
\be
	M = \frac{\beta_l}{4 \pi}\ , \, \, \, Q = \frac{\varphi_l}{2 \pi}\ , \, \, \, P = \frac{\xi_l}{2 \pi}\ ,
\ee
\bea
\label{eq:asympttor}
\begin{split}
	& M =  \sqrt{\frac{\beta_r^2 - 4 \varphi_r^2 + \beta_r \sqrt{\beta_r^2 + 8 \varphi_r^2}}{8 \pi^2 + 2 \omega_r^2}} \ , \\
	Q = & \frac{-\beta_r + \sqrt{\beta_r^2 + 8 \varphi_r^2}}{4 \varphi_r}\, M\ , \quad a = \frac{3 \beta_r - \sqrt{\beta_r^2 + 8 \varphi_r^2}}{4 (4 \pi^2 + \omega_r^2)}\, \omega_r\ ,
\end{split}
\eea
where we again suppressed $P$ and $\xi_r$ for brevity. Using these, we finally arrive at one of the central results in this work, the explicit expressions in \eqref{eq:expl}. The mixed ensemble instead gives
\be
	\tilde I_{l,r} = I_{l,r} + \xi_{l,r} P, \qquad \frac{\partial I_{l,r}}{\partial \xi_{l,r}} = - P\ ,
\ee
such that in the left moving sector one finds
\be
\label{eq:mixedl}
	\tilde I_l (\beta_l, \varphi_l, P) = \frac1{8 \pi}\, (\beta_l^2 - 2 \varphi_l^2) + \pi\, P^2\ .
\ee
$\tilde I_r(\beta_r, \omega_r, \varphi_r, P)$ follows straightforwardly in the same way, but the expression is much more lengthy and we spare it from the reader here.

\vspace{-5mm}
\subsubsection{Conjugate variables and stability}
\label{ssubec:conj}
\vspace{-3mm}
Note an apparent asymmetry in the presentation so far, since we have substituted the variables $\Omega_\pm, \Phi_\pm, \Psi_\pm$ for $\omega_{l,r}, \varphi_{l,r}, \xi_{l,r}$. The fact that these are the canonical variables in their respective ensembles is determined by the following set of identities. For the standard on-shell actions $I_\pm$ (or $\tilde I_\pm$), in the limits when it is possible to check (see below), one can verify
\bea
\begin{split}
	\frac{\partial I_\pm}{\partial \beta_\pm} &= M - \Omega_\pm J - \Phi_\pm Q\ , \\
\frac{\partial I_\pm}{\partial \Omega_\pm} &= - \beta_\pm J\ , \quad \quad \frac{\partial I_\pm}{\partial \Phi_\pm} = - \beta_\pm Q\ ,
\end{split}
\eea
which is expected from the direct relation with the Gibbs free energies, $G_\pm = I_\pm / \beta_\pm$, and justifies us writing the quantum statistical relation in the form of \eqref{eq:qsr}. Instead, for the new on-shell actions $I_{l,r}$ (or $\tilde I_{l,r}$) we find
\be
		\frac{\partial I_{l,r}}{\partial \beta_{l,r}} = M \ , \quad \frac{\partial I_{r}}{\partial \omega_{r}} = - J\ , \quad \frac{\partial I_{l,r}}{\partial \varphi_{l,r}} = - Q\ ,
\ee
which is the reason we used the conjugate variables $\omega, \varphi$ in \eqref{eq:qsrnew} and elsewhere. 

Using the explicit formulae above, \eqref{eq:lrinasympt}-\eqref{eq:asympttor}, it is straightforward to compute the heat capacities  \cite{Hristovtoapp} and look at the criteria for local and global thermodynamic stability. We should however note that  in the absence of quantum corrections this analysis is qualitatively inconclusive  in standard variables already for the outer horizon, see e.g.\ \cite{Cvetic:2018dqf,Avramov:2023eif}. Perhaps a clearer sign of stability in this setting is the semi-classical calculation of \cite{Cvetic:1997uw,Cvetic:1997xv}, where it was shown that scalar perturbations in the vicinity of the two horizons precisely feel a thermal bath of the corresponding combination of $\beta_l$ and $\beta_r$. Furthermore, the inner horizon was shown to be a regular Cauchy horizon, \cite{Ansorg:2009yi}, sparking much interest in the area product formula. Consequently, such a formula was shown to hold for black objects with different asymptotics in various dimensions, see e.g.\ \cite{Cvetic:2010mn,Castro:2012av}. Although the area product in itself is not immediately useful here for evaluating the on-shell actions, the related literature strongly suggests that the use of left and right moving variables is widely applicable.

\vspace{-5mm}
\subsection{Notable limits}
\label{sec:limits}
\vspace{-3mm}

There are several limits which do allow for an explicit form of the on-shell action already in terms of the standard chemical potentials. In the Reissner-Nordstr\"om limit, $a = J = \Omega_\pm = 0$, one can verify that
\be
	I^\text{RN}_\pm = \frac{\beta_\pm^2 \left(1 - \Phi_\pm^2 - \Psi_\pm^2 \right)^2}{16 \pi}\ ,
\ee
while in the Kerr limit, $Q = P = \Phi_\pm = \Psi_\pm = 0$, one finds
\be
	I^\text{Kerr}_\pm = \frac{1}{4 \Omega_\pm^2}\, \left( \pm \sqrt{4 \pi^2 + \beta_\pm^2 \Omega_\pm^2} - 2 \pi\right)\ .
\ee
Both limits of $I_+$ in turn lead to the same Schwarzschild limit where only $M \neq 0, \beta_+ \neq 0$ (for a single horizon quantities with ``$-$'' subscript are not defined), 
\be
	I^\text{Schw}_+ = \frac{\beta_+^2}{16 \pi}\ , \quad \quad I_- = 0\ .
\ee

In terms of the new variables, we do not find a particular simplification of \eqref{eq:expl} in the Reissner-Nordstr\"om limit, $ \omega_r = 0$. In the Kerr limit, $\varphi_\pm = \xi_\pm = 0$, one finds
\be
	I^\text{Kerr}_l = \frac{\beta_l^2}{8 \pi}\ , \quad \quad I^\text{Kerr}_r = \frac{\beta_r^2}{4 \sqrt{4 \pi^2 + \omega_r^2}}\ ,
\ee
leading to the expected Schwarzschild limit $\beta_l = \beta_r = \beta$,
\be
	I^\text{Schw}_l = I^\text{Schw}_r = \frac{\beta^2}{8 \pi}\ .
\ee

\vspace{-5mm}
\subsubsection{BPS and extremal limits}
\label{ssubec:bps}
\vspace{-3mm}
In the BPS limit, $M = \sqrt{Q^2 + P^2}$ while keeping $J \neq 0$, we could not find a simple expression for the on-shell action in terms of the standard chemical potentials. In \cite{Hristov:2022pmo} it was shown that a working recipe is to shift $\Phi_\pm$ and $\Psi_\pm$ by their extremal values, which conversely does not seem useful away from this limit. Remarkably, we find that in terms of the new variables we formally reproduce the exact same result as in \cite{Hristov:2022pmo}. The thermodynamic potentials are now manifestly {\it complex} and further satisfy
\be
	\beta_l = 2\, \sqrt{\varphi_l^2 + \xi_l^2}\ , \, \, \omega_r = \pm 2 \pi \i\ , \, \, \beta_r = \sqrt{\varphi_r^2 + \xi_r^2}\ ,
\ee
such that we arrive at the anticipated
\be
	I^\text{BPS}_l  =  \frac{\varphi_l^2 + \xi_l^2}{4 \pi}\ , \quad \quad \quad I^\text{BPS}_r = 0\ .
\ee
The on-shell action no longer depends on the inverse temperature, as expected given that the mass is no longer free. The first law of thermodynamics in the BPS limit, leading to the corresponding quantum statistical relation, is given by
\be
	\delta I^\text{BPS}_l = - \delta S_l + \varphi_l\, \delta Q + \xi_l\, \delta P = 0 \ ,
\ee
with opposite signs for $\varphi_l, \xi_l$ compared to the thermal case, \eqref{eq:lrfirstlaw}, due to the additional contribution from the $\beta_l\, \delta M$ term that has been absorbed inside,~\footnote{Therefore the precise map to the BPS limit in \cite{Hristov:2022pmo} involves $\varphi_l, \xi_l$ taken with a negative sign. In presence of scalars this map is more involved, \cite{Hristovtoapp}, but remains linear.}. Curiously, this leads to a mixed ensemble expression
\be
\label{eq:mixedbps}
	\tilde I^\text{BPS}_l (\varphi_l, P)  = I^\text{BPS}_l  - \xi_l P =   \frac{\varphi_l^2}{4 \pi} - \pi P^2\ ,
\ee 
which is clearly different from the direct BPS limit of $\tilde{I}_l (\beta_l, \varphi_l, P)$, \eqref{eq:mixedl}. 

The extremal limit, which is a priori given by $M =\sqrt{a^2 + Q^2 + P^2}$ is not directly well-defined in the right moving sector. However, one can reach it in steps by first taking the BPS limit and then taking $a = J = 0$, which clearly just reproduces the formulae above as they are independent of $J$. In this case the mixed ensemble evaluation can be done in a direct manner in Sen's entropy function approach, \cite{Sen:2005wa}, by focusing on the near-horizon geometry exhibiting an AdS$_2$ factor. The evaluation of the near-horizon on-shell action produces precisely \eqref{eq:mixedbps}. The fact that the on-shell action is purely holomorphic is instead important for the OSV conjecture in the presence of higher derivative corrections, \cite{Ooguri:2004zv}.

\vspace{-5mm}
\subsection{New thermal partition function}
\label{sec:pf}
 \vspace{-3mm}
Let us now discuss the quantum gravity implications of our results. Assuming a finite UV completion of the Einstein-Maxwell theory, in the standard interpretation of \cite{Gibbons:1976ue} it is natural to consider the thermal partition function in the grand-canonical ensemble $Z (\beta_+, \Omega_+, \Phi_+)$,~\footnote{For simplicity of the presentation we are again suppressing the magnetic charges and discuss the thermodynamics only on the outer horizons as in the original references.}, which in a microscopic theory should follow from the path integral formalism. No matter what the full answer is, at a semi-classical level (denoted here with $\approx$) we have
\be
	Z^\text{gc} (\beta_+, \Omega_+, \Phi_+) \approx \exp[ - I_+ (\beta_+, \Omega_+, \Phi_+) ]\ ,
\ee
such that the microcanonical thermal partition function is found via an inverse Laplace transform,
\be
	Z^\text{mc} (M, J, Q) = \int [{\rm d} \phi_+]\, Z^\text{gc}\, e^{ \beta_+ M - \beta_+ \Omega_+ J - \beta_+ \Phi_+ Q}\ ,
\ee
where $[{\rm d} \phi_+] := {\rm d} \beta_+ {\rm d} \Omega_+  {\rm d} \Phi_+$. In the saddle point approximation we recover the Bekenstein-Hawking entropy,
\be
	\log Z^\text{mc} \approx S_+\ ,
\ee
and the full answer for $Z^\text{mc}$ is the exact quantum degeneracy of the black hole. Our results suggest the existence of a new set of partition functions based on the holomorphic and anti-holomorphic variables instead. The grand-canonical version should be such that
\be
	Z^\text{gc}_{l,r} (\beta_{l,r}, \omega_{l,r}, \varphi_{l,r}) \approx \exp[- I_{l,r} (\beta_{l,r}, \omega_{l,r}, \varphi_{l,r}) ]\ ,
\ee
giving us the opportunity to define a genuinely new microcanonical partition function, $\hat Z^\text{mc}$,
\bea
\label{eq:newpf}
\begin{split}
	\hat Z^\text{mc} (M, J, Q) := &\int [{\rm d} \phi_l]\, Z^\text{gc}_l\, e^{\beta_l M - \varphi_l Q} \\
& \times    \int [{\rm d} \phi_r]\, Z^\text{gc}_r\, e^{ \beta_r M - \omega_r J - \varphi_r Q}\ ,
\end{split}
\eea
which by construction also satisfies
\be
	\log \hat Z^\text{mc} \approx S_l + S_r = S_+\ .
\ee
Due to the different integration variables (and corresponding integration contours) it is clear that the two microcanonical partition functions are a priori completely independent. Notice in particular the similarity between \eqref{eq:newpf} and the form of the thermal partition function for 2d CFT's, see e.g. Section 2 in \cite{Hosseini:2020vgl}. This analogy was also pursued in \cite{Strominger:1997eq} and is again based on the introduction of the left and right moving sectors for the BTZ black hole. From a microscopic point of view there is no obvious reason why the left moving and right moving sectors should have the same conserved charges as is the case above, $\{M_l, Q_l \} = \{ M_r, Q_r \} = \{ M, Q \}$, but this could well be a leading order gravitational effect similar to the equality $c_l = c_r$ in the BTZ case, \cite{Kraus:2005zm}.

\vspace{-5mm}
\subsection{Outlook}
\label{sec:out}
\vspace{-3mm}
The present results are imminently generalizable \cite{Hristovtoapp} to other 4d asymptotically flat black holes in extensions of Einstein-Maxwell theory with additional scalars and Maxwell fields such as STU black holes, \cite{Chow:2014cca}. More importantly, the new set of variables can be defined on any pair of black hole horizons away from extremality, i.e.\ for {\it every} thermal black hole background, their wide applicability already suggested by \cite{Cvetic:2010mn,Castro:2012av}, see also \cite{Guica:2008mu} and references thereof. However, it would be naive to expect that one can always solve the ``explicitness'' problem since the horizon radii are often given by more complicated expressions, e.g.\ coming from quartic instead of quadratic equations. Nevertheless, the argued naturalness of the new variables is expected to hold in many other interesting examples, which we hope to explore in future. It would be interesting to find an overarching pattern in the evaluation of black hole on-shell actions  in the spirit of \cite{Gibbons:1979xm,Hosseini:2019iad} but now also away from the BPS/extremal limit.

\vspace{-5mm}
\subsubsection*{Acknowledgements}
\vspace{-3mm}
I am very grateful to S.\ M.\ Hosseini, C.\ Toldo and S.\ Yazadjiev for helpful discussions. I am supported in part by the Bulgarian NSF grant KP-06-N68/3.

\bibliographystyle{apsrev4-2}
\bibliography{newthermo.bib}

\end{document}